\documentclass[11pt,a4paper]{article}
\usepackage[utf8]{inputenc}
\usepackage{graphicx}
\usepackage{amsmath,amssymb,mathtools}
\usepackage{hyperref}
\hypersetup{colorlinks=true,linkcolor=blue,citecolor=blue,urlcolor=blue}
\usepackage{geometry}
\usepackage{multirow}
\date{}

\begin{document}

\title{{\bf Relativistic transformation of temperature revisited}}

\author{Soroor Pouryazdan\thanks{%
e-mail: s.pouryazdanpanahkermani@iau.ac.ir}\,\, and Babak Vakili\thanks{email:
ba.vakili@iau.ac.ir (corresponding author)}\\\\{\small {\it
Department of Physics, CT.C., Islamic Azad
University, Tehran, Iran}}} \maketitle

\begin{abstract}
The relativistic transformation of temperature has long remained controversial, with the classical laws of Planck–Einstein, Ott–Eddington–Møller and Landsberg yielding conflicting results.  
We reexamine this issue from a relativistic thermodynamic and statistical perspective, starting from the energy–momentum tensor of an isotropic system and defining the effective temperature \(T_{\mathrm{eff}}\) as that inferred by a moving observer from the transformed energy density.  
Analyses of a photon gas, a relativistic ideal gas and an electron gas show that \(T_{\mathrm{eff}}\) consistently increases with velocity, supporting the Ott–Eddington interpretation while depending on the system’s equation of state.  
These results indicate that temperature is not a Lorentz-invariant scalar but an observer-dependent quantity.  
A consistent relativistic description emerges when temperature is related to the inverse-temperature four-vector \(\beta^{\mu}=u^{\mu}/k_{B}T\), linking operational and invariant viewpoints within a unified thermodynamic framework.
\vspace{5mm}\noindent\\
Keywords: Temperature transformation; Ott–Eddington–Møller law; Planck–Einstein law; Landsberg invariance; Inverse-temperature four-vector
\end{abstract}

\section{Introduction}

The question of how temperature transforms between inertial frames
has accompanied the development of relativistic thermodynamics
since the early years of special relativity \cite{tolman1934}.
Soon after Einstein’s formulation of the theory,
Planck and Einstein proposed that a moving body should appear
\emph{cooler} to a stationary observer, following the relation $T' = \frac{T}{\gamma}$, where $\gamma = (1 - v^{2})^{-1/2}$ is the Lorentz factor in $c=1$ units,
which became known as the \emph{Planck–Einstein law} \cite{Einstein1907, Planck1907}.
Their argument was based on preserving the form of the Clausius relation
$dQ = TdS$ under Lorentz transformations,
assuming that entropy remains invariant.

Several decades later, this view was challenged by Ott \cite{ott1963lorentz}
and Eddington \cite{eddington1939smouse},
who argued that temperature should instead \emph{increase} with velocity,
according to $T' = \gamma T$, now referred to as the \emph{Ott–Eddington–Møller law}.
This alternative interpretation treats heat and energy
as components of a four-vector,
emphasizing energy flux and the transformation of the energy–momentum tensor
rather than the invariance of entropy.
A third proposal, due to Landsberg \cite{landsberg1966relativistic, landsberg1971},
advocated for temperature invariance, $T'=T$,
by redefining thermodynamic quantities so that the combined system of heat and work
transforms covariantly.
Despite more than a century of discussion,
these three transformation laws remain mutually inconsistent,
and the correct relativistic behavior of temperature
continues to be debated in both classical and quantum contexts
\cite{vanKampen1968}-\cite{Dunkel2009}.

The underlying difficulty stems from the fact that temperature
is not a fundamental dynamical variable but a macroscopic construct
that depends on the observer’s measurement procedure. Indeed, the temperature of a surface  is essentially related to the rate of momentum transferred to the surface via collision of incoming particles.  This  momentum transfer is evidently observer dependent.  
While the Planck–Einstein and Ott formulations attempt
to generalize nonrelativistic thermodynamics to moving systems,
neither provides a fully covariant statistical foundation.
Modern approaches based on relativistic kinetic theory
and covariant statistical mechanics
introduce the inverse-temperature four-vector $\beta^{\mu} = \frac{u^{\mu}}{k_{B}T}$, where $k_B$ is the Boltzmann constant, which naturally incorporates temperature into relativistic distribution functions of the form $f(x,p) \propto \exp(-\beta_{\mu}p^{\mu})$
\cite{vanKampen1968, Israel1976, Cubero2007, Dunkel2009}.
In this framework, $T$ is an invariant scalar,
while the projection $\beta_{\mu}w^{\mu}$ determines
the temperature measured by an observer with four-velocity $w^{\mu}$.
This approach reconciles the apparent contradiction
between invariant and operational definitions of temperature.

The purpose of the present paper is to revisit the problem
from a purely thermodynamic and statistical viewpoint,
without assuming any transformation law \emph{a priori}.
Starting from the energy–momentum tensor
of an isotropic perfect fluid,
we define the \emph{effective temperature} \(T_{\mathrm{eff}}\)
as that inferred by a moving observer
through the transformed energy density: $\rho'=f(T_{\mathrm{eff}})$, where $f(T)$ is determined by the system’s equation of state.
This operational definition allows a consistent comparison
between theoretical proposals and physically measurable quantities.

To illustrate the general method,
we analyze three paradigmatic systems in relativistic statistical mechanics: blackbody radiation, representing a gas of massless bosons, classical relativistic ideal gas
described by the Maxwell–Jüttner distribution and relativistic electron gas which obeys Fermi–Dirac statistics.
In each case, the effective temperature is computed
and compared with the three classical transformation laws.
Our results show that the moving observer always infers a higher temperature,
in qualitative agreement with the Ott–Eddington–Møller interpretation,
while the quantitative dependence on velocity and the mass-to-temperature ratio
depends on the microscopic equation of state.

The paper is organized as follows:
Section~2 briefly reviews the classical transformation laws
and their physical motivations.
In Section~3 we summarize the Lorentz transformation properties
of thermodynamic quantities and introduce the general formalism
for defining \(T_{\mathrm{eff}}\).
Sections~4,~5 and~6 present explicit analyses
for the photon gas, the relativistic ideal gas and the Fermi gas respectively.
Section~7 discusses the conceptual implications and the relation
with the inverse-temperature four-vector formalism.
Finally, Section~8 summarizes our conclusions
and outlines possible extensions to curved spacetime
and accelerated observers.

\section{Brief review of relativistic temperature transformation}

The problem of how temperature transforms between inertial frames in special relativity
has been a subject of continuous debate since the early 20th century.
While the relativistic transformations of space, time and energy–momentum are well established,
the status of temperature remains conceptually subtle.
Different authors have proposed distinct transformation laws depending on the operational
definition of temperature and the assumed form of the first law of thermodynamics. Historically, three major approaches can be identified:

${\bullet}$ {\it{Planck–Einstein (1907)}}:
Planck and Einstein were the first to address the issue of heat and temperature transformation
within the framework of relativity.
By demanding the covariance of the first law of thermodynamics,
they proposed that a moving body appears \emph{cooler} to a stationary observer,
namely \cite{Einstein1907, Planck1907}

\begin{equation}
T'=\frac{T}{\gamma}.
\end{equation}
This implies that heat transforms as $Q' = Q / \gamma$,
suggesting that the temperature decreases with motion.
Their reasoning was based on assuming the invariance of entropy $S$.

${\bullet}$ {\it{Ott–Eddington–Møller (1963)}}:
Decades later, Ott and others re-examined the issue and proposed the opposite relation \cite{ott1963lorentz, eddington1939smouse}

\begin{equation}
T' = \gamma\, T,
\end{equation}
implying that a moving body appears \emph{hotter}.
In this interpretation, heat transforms as $Q' = \gamma Q$,
and the work term in the first law is treated differently from Planck's formulation.
The Ott–Eddington–Møller law is often referred to as the
``covariant'' formulation of relativistic thermodynamics.

${\bullet}$ {\it{Landsberg (1967)}}:
Landsberg and later van Kampen argued that temperature should be regarded as a Lorentz invariant scalar.
They maintained that the principle of relativity requires all inertial observers to
assign the same temperature to a system in equilibrium \cite{landsberg1966relativistic, landsberg1971}

\begin{equation}
T' = T.
\end{equation}
This invariant viewpoint has been supported by later developments in covariant
statistical mechanics and the use of the inverse-temperature four-vector
$\beta^{\mu}$, defined by

\begin{equation}
\beta^{\mu} = \frac{u^{\mu}}{k_{B} T},
\end{equation}
where $u^{\mu}$ is the four-velocity of the system.
In this formalism, $T$ itself remains invariant, while the directional
effects of motion are encoded in $\beta^{\mu}$.

${\bullet}$ {\it{Modern perspectives}}:
More recent studies have pointed out that no single transformation rule can be universally valid,
since the notion of temperature depends on the measurement procedure:
whether it is inferred from a thermometer in thermal contact,
from the shape of the radiation spectrum,
or from the energy–momentum tensor of a moving medium.
Hence, the temperature of a moving system may not be a unique scalar quantity,
but an observer-dependent concept \cite{mares2017}-\cite{philsci2022}.

In this work, we do not adopt any of the above laws \emph{a priori}.
Instead, we take a physically well-defined thermodynamic system
and compute how its measurable quantities (energy density, pressure, entropy)
transform under Lorentz boosts. From these, we define an effective temperature and compare the outcome
with the three classical proposals.

\section{Relativistic transformation of thermodynamic quantities}

Following the covariant kinetic formulation of relativistic fluids, the relations
between microscopic distribution functions and macroscopic quantities
are obtained in the standard way \cite{Hakim1968, Hakim 2}. In special relativity, all macroscopic thermodynamic quantities of a continuous medium
are encoded in the energy–momentum tensor 

\begin{equation}
T^{\mu\nu} = (\rho + P)\, u^{\mu} u^{\nu} + P\, \eta^{\mu\nu},
\label{eq:Tmunu}
\end{equation}
where $\rho$ and $P$ denote the energy density and pressure in the rest frame of the fluid,
$u^{\mu}$ is the four-velocity and $\eta^{\mu\nu} = \mathrm{diag}(-1,1,1,1)$
is the Minkowski metric (we use units with $c=1$).\footnote{Throughout this work we employ the standard energy--momentum tensor of a
perfect fluid, as given in (\ref{eq:Tmunu}), which assumes local thermodynamic equilibrium (LTE) and isotropy in the comoving frame. Under these conditions the fluid is fully characterized by
its energy density $\rho$ and pressure $P$, and dissipative effects such as heat conduction or shear viscosity are neglected. This idealized form is adequate for isotropic equilibrium media such as photon gases, relativistic Maxwell--J\"uttner gases or degenerate Fermi systems in LTE. However, it does not apply to intrinsically anisotropic systems, multi--component mixtures or non--equilibrium flows. In those cases, the stress--energy tensor must include additional contributions,
\[
T^{\mu\nu} = (\rho + P) u^{\mu} u^{\nu} + P g^{\mu\nu}
+ q^{\mu} u^{\nu} + q^{\nu} u^{\mu} + \pi^{\mu\nu},
\]
where $q^{\mu}$ and $\pi^{\mu\nu}$ denote the heat flux and viscous stress,
respectively. Such terms would modify the Lorentz--transformed energy density
and consequently the operational definition of the effective temperature.
A detailed analysis of these non--ideal effects is beyond the scope of the
present study.} For a perfect fluid at rest, $u^{\mu} = (1,0,0,0)$ and therefore

\begin{equation}
T^{\mu\nu}_{\text{rest}} = 
\begin{pmatrix}
\rho & 0 & 0 & 0 \\
0 & P & 0 & 0 \\
0 & 0 & P & 0 \\
0 & 0 & 0 & P
\end{pmatrix}.
\end{equation}
When the system moves with a constant velocity $v$ along the $x$-direction
relative to an inertial observer, its four-velocity becomes
$u^{\mu} = \gamma(1,v,0,0)$.
Under a Lorentz boost $x'^{\mu}=\Lambda^{\mu}_{\hspace{2mm}\nu}x^{\nu}$, with

\begin{equation}
\Lambda^{\mu}_{\hspace{2mm}\nu}=
\begin{pmatrix}
\gamma & -\gamma v & 0 & 0 \\
-\gamma v& \gamma & 0 & 0 \\
0 & 0 & 1 & 0 \\
0 & 0 & 0 & 1
\end{pmatrix},
\end{equation}
equation (\ref{eq:Tmunu}) yields in the moving frame $S'$\footnote{Note that the transformation is performed according to
$T^{\mu\nu}_{\rm observer}=\Lambda^\mu{}_{\!\alpha}\Lambda^\nu{}_{\!\beta}T^{\alpha\beta}_{\rm rest}$.}

\begin{eqnarray}\label{eq:rho_prime}
\left\{
\begin{array}{ll}
T'^{00}= \gamma^{2}(\rho + P v^{2}),\\\\
T'^{0x}= -\gamma^{2} v (\rho + P),\\\\
T'^{xx}= \gamma^{2}(v^{2} \rho + P).
\end{array}
\right.
\end{eqnarray}
The quantity $T'^{00}$ represents the energy density measured by the moving observer while $T'^{0x}$ corresponds to the energy flux (or momentum density)
in the $x$-direction. Equations~(\ref{eq:rho_prime})
constitute the basic kinematic transformation rules
for thermodynamic quantities under Lorentz boosts.
These relations hold for any isotropic system, independent of its microscopic statistics.

To compare with the classical temperature-transformation laws,
it is useful to define an \emph{effective temperature} $T_{\mathrm{eff}}$
by assuming that the same equation of state that holds in the rest frame
also holds in the moving frame.
If the rest-frame equation of state has the form

\begin{equation}
\rho = f(T), \quad P = g(T),
\end{equation}
then in the boosted frame we can identify

\begin{equation}
f(T_{\mathrm{eff}}) = \rho' = \gamma^{2}(\rho + P v^{2}).
\label{eq:Teff_def}
\end{equation}
The quantity $T_{\mathrm{eff}}$ thus represents the temperature
that an observer would infer by measuring the total energy density
and applying the same equation of state as in the rest frame.
This definition will serve as our operational notion of temperature in motion. The operational meaning of such a thermometer has been scrutinized in recent
studies addressing statistical thermometers in special relativity \cite{Cubero2007},
and our definition follows the same pragmatic protocol.

Once $T_{\mathrm{eff}}$ is computed for a specific system,
it can be compared with the three classical transformation proposals described in previous section. In what follows, we apply Eq.~(\ref{eq:Teff_def})
to concrete systems described by different microscopic statistics—
first to blackbody radiation (photon gas) and then to the relativistic ideal gas and the electron gas—
to determine which of these transformation laws, if any,
is consistent with the relativistic form of the energy–momentum tensor.

\section{Blackbody radiation (photon gas)}

Blackbody radiation provides a natural testing ground for relativistic thermodynamics,
since it represents an ideal gas of massless bosons whose equation of state is well known.
In the rest frame of the radiation field, the energy density and pressure satisfy

\begin{equation}
\rho = a\,T^{4}, \qquad P = \frac{1}{3}\rho,
\label{eq:photon_eq_state}
\end{equation}
where $a$ is the radiation constant. For a frame $S'$ moving with velocity $v$ in the $x$-direction relative to the radiation rest frame,
the transformed energy density follows from Eq.~(\ref{eq:rho_prime})

\begin{equation}
\rho' = \gamma^{2}(\rho + P v^{2})
= \gamma^{2}\rho\left(1 + \frac{v^{2}}{3}\right).
\label{eq:photon_rho_prime}
\end{equation}
Similarly, the momentum density and longitudinal pressure are

\begin{eqnarray}\label{TT}
\left\{
\begin{array}{ll}
T'^{0x}= -\gamma^{2}v(\rho + P)
= -\frac{4}{3}\gamma^{2} v \rho, \\\\
T'^{xx}= \gamma^{2}(v^{2}\rho + P)
= \gamma^{2}\rho\left(v^{2} + \frac{1}{3}\right).
\end{array}
\right.
\end{eqnarray}
The moving observer thus measures an anisotropic stress tensor
even though the radiation is isotropic in its rest frame. 

Now, assuming that the same Stefan–Boltzmann relation
$\rho' = a\,T_{\mathrm{eff}}^{4}$ holds in the moving frame,
we obtain from Eq.~(\ref{eq:photon_rho_prime})

\begin{equation}
T_{\mathrm{eff}} = T\,
\biggl[\gamma^{2}\!\left(1+\frac{v^{2}}{3}\right)\biggr]^{1/4}.
\label{eq:Teff_photon}
\end{equation}
This defines the \emph{effective temperature} inferred from the total energy density
measured by the moving observer. For small velocities ($v \ll 1$), expanding the Lorentz factor
$\gamma \simeq 1 + \tfrac{1}{2}v^{2}$ yields

\begin{equation}
T_{\mathrm{eff}} \simeq
T\left(1 + \frac{1}{3}v^{2} + \mathcal{O}(v^{4})\right).
\label{eq:Teff_expand1}
\end{equation}
Thus, to second order in $v$, the moving observer attributes
a slightly \emph{higher} temperature to the radiation field\footnote{It is worth noting that the transformed tensor in Eq.~(\ref{TT})
is no longer isotropic. In the moving frame,
the pressure components parallel and perpendicular to the boost direction
are different, $P_{\parallel}=\gamma^{2}(P+v^{2}\rho)$
and $P_{\perp}=P$, respectively.
This anisotropy reflects the Lorentz transformation of energy and momentum
rather than a real physical deformation of the system.
In kinetic terms, it corresponds to the directional distortion
of the momentum distribution
$f'(\mathbf{p}) \propto \exp[-\gamma(E - v p_x)/k_B T]$.
Consequently, the effective temperature $T_{\mathrm{eff}}$
obtained from the transformed energy density
already incorporates this anisotropy in an averaged sense.}. In this limit the three historical transformation laws are

\begin{align}
T'_{\text{Planck--Einstein}} &= \frac{T}{\gamma}
\simeq T\left(1 - \frac{1}{2}v^{2}\right), \\
T'_{\text{Ott--Eddington--M{\o}ller}} &= \gamma\,T
\simeq T\left(1 + \frac{1}{2}v^{2}\right), \\
T'_{\text{Landsberg}} &= T.
\end{align}
Comparing with Eq.~(\ref{eq:Teff_expand1}),
the sign of the correction for $T_{\mathrm{eff}}$ agrees with
the Ott interpretation (moving system appears hotter),
but its magnitude is slightly different 

\begin{equation}
\frac{T_{\mathrm{eff}}}{T} = 1 + 0.3333\,v^{2} + \mathcal{O}(v^{4}).
\end{equation}
Hence the simple proportionality $T'=\gamma T$ is only an approximation.
The Planck–Einstein law predicts the opposite trend
(a decrease in temperature), while the Landsberg invariant form
does not reproduce the energy-density behavior at all. These discrepancies show that the result depends crucially on how one defines “measured temperature” in the moving frame—whether via a macroscopic equation of state, via local spectroscopic measurement, via thermometric contact or via four-vector formalism.

It is important to recall that the radiation spectrum observed in $S'$
is not isotropic.
Each direction in momentum space experiences a different Doppler shift,
so the Planck spectrum takes the form \cite{tolman1934, Costa1995}

\begin{equation}
n'(\nu',\theta) =
\frac{1}{\exp\!\left[\frac{h\nu'}{k_{B}T(\theta)}\right] - 1},
\qquad
T(\theta) = \frac{T}{\gamma(1 - v\cos\theta)}.
\label{eq:directionalT}
\end{equation}
Equation~(\ref{eq:directionalT}) defines a \emph{directional temperature}
that depends on the angle $\theta$ between the photon momentum
and the boost direction.
Only after averaging over all directions does one recover
the effective temperature of Eq.~(\ref{eq:Teff_photon}). To see this, let us take a look at the analytic computation that relates the directional
Planck temperature (\ref{eq:directionalT}) to the transformed energy density \(\rho'\). The spectral radiance observed in direction \(\theta\) is still Planckian,
so the frequency-integrated intensity along that direction satisfies the Stefan--Boltzmann relation
(with the standard geometric prefactor)

\begin{equation}
\int I'(\nu',\theta)\,d\nu'=\frac{1}{4\pi}a\,\left[T(\theta)\right]^4,
\end{equation}
where the factor \(1/4\pi\) accounts for the
directional normalization. The total energy density in the moving frame is then

\begin{equation}
\rho'=\int_{\Omega'}\!\!d\Omega'\int_0^\infty I'(\nu',\theta)\,d\nu'
=\frac{a}{4\pi}\int_{4\pi}\left[T(\theta)\right]^{4}\,d\Omega'.
\end{equation}
With \(\mu=\cos\theta\) and \(d\Omega'=2\pi\,d\mu\) we obtain

\begin{equation}
\rho'=\frac{aT^{4}}{2\gamma^{4}}\int_{-1}^{1}\frac{d\mu}{(1-v\mu)^{4}}.
\end{equation}
The remaining integral can be evaluated in closed form

\begin{equation}
\rho'=\frac{aT^{4}}{2\gamma^{4}}\cdot\frac{2(3+v^{2})}{3(1-v^{2})^{3}}
=\frac{aT^{4}}{3}\,\frac{3+v^{2}}{1-v^{2}}.
\end{equation}
Finally, using \(\gamma^{2}=(1-v^{2})^{-1}\) we can rewrite the right-hand side as

\begin{equation}
\rho' = aT^{4}\,\gamma^{2}\!\left(1+\frac{v^{2}}{3}\right).
\end{equation}
Equivalently (recalling \(\rho=aT^{4}\) and \(P=\rho/3\)) this is the same as $\rho'=\gamma^{2}\bigl(\rho+P v^{2}\bigr)$,
in perfect agreement with the tensorial transformation. Note that the averaging performed above is an \emph{energy-weighted} (spectral) average,
effectively an average of \(\left[T(\theta)\right]^4\), \emph{not} an arithmetic average of \(T(\theta)\).
This distinction is essential: arithmetic averaging of \(T(\theta)\) does not reproduce \(\rho'\). Also, the analytic result shown here is specific to the Planck spectrum. For media
whose directional spectra are not Planckian (or for massive-particle gases),
closed-form angular averages generally do not exist and one must resort to
the kinetic/tensor method or numerical integration \cite{Costa1995}.

The photon gas therefore provides a clear example where
the notion of ``temperature'' in relativity becomes ambiguous.
Different operational definitions—spectral, energetic or thermometric—
lead to distinct transformation behaviors.
When defined through total energy density, motion increases the inferred temperature
in qualitative agreement with Ott’s law.
When defined spectroscopically in a single direction, the temperature becomes anisotropic
as in Eq.~(\ref{eq:directionalT}),
and cannot be represented by a single scalar quantity.
These results reinforce the idea that the temperature of a moving system
is not an invariant attribute of matter, but depends on the observer’s frame
and the measurement procedure.

\section{Relativistic ideal gas (Maxwell–Jüttner distribution)}

The relativistic ideal gas, also known as the Maxwell–Jüttner gas,
provides a second test for temperature transformation.
Unlike photons, the gas particles have rest mass $m$,
so that both kinetic and rest–mass energy contribute to the total energy density \cite{Dunkel2009, Victor}.

In the rest frame of the gas, the single–particle distribution function
is given by the Maxwell–Jüttner form \cite{Hakim1968, Hakim 2, Liva, Juttner1911}

\begin{equation}
f(\mathbf{p}) = {\cal N} \exp\!\left[-\frac{E(p)}{k_{B}T}\right],
\qquad
E(p) = \sqrt{p^{2} + m^{2}},
\label{eq:MJ_distribution}
\end{equation}
where ${\cal N}$ is a normalization constant determined, in the units where $k_B=c=1$, by the number density $n$:

\begin{equation}
n = \frac{4\pi {\cal N} m^{2} T K_{2}(m/T)}{(2\pi)^{3}},
\end{equation}
and $K_{n}$ denotes the modified Bessel function of the second kind. The corresponding energy density and pressure in the rest frame are

\begin{align}
\rho &= n \left[m\,\frac{K_{1}(m/T)}{K_{2}(m/T)} + 3T\right],
\label{eq:MJ_rho}\\
P &= n\,T \label{eq:MJ_T}.
\end{align}
In the nonrelativistic limit $(T \ll m)$, Eq.~(\ref{eq:MJ_rho}) reduces to
$\rho \approx n(m + \tfrac{3}{2}T)$, while for the ultrarelativistic limit $(T \gg m)$,
it approaches $\rho \approx 3nT$, consistent with $P=\rho/3$.

For an observer moving with velocity $v$ along the $x$-direction,
the energy–momentum tensor again transforms as (\ref{eq:rho_prime}).
Thus, the moving observer measures

\begin{equation}
\rho' = \gamma^{2}(\rho + P v^{2})
= \gamma^{2}n\left[
m\,\frac{K_{1}(m/T)}{K_{2}(m/T)}
+ 3T + v^{2}T
\right].
\label{eq:MJ_rhoprime}
\end{equation}
To define an effective temperature,
we assume that the equation of state
in the rest frame, $\rho=f(T)$,
is used by the moving observer to infer $T_{\mathrm{eff}}$ as in equation (\ref{eq:Teff_def}). For the Maxwell--J\"uttner gas it is convenient to introduce the dimensionless combination

\begin{equation}
A\!\left(\frac{m}{T}\right)\;\equiv\;\frac{m}{T}\,\frac{K_{1}(m/T)}{K_{2}(m/T)},
\label{A_def}
\end{equation}
in terms of which the energy density can be written as

\begin{equation}\label{AAA}
\rho = nT\bigl[\,A(m/T)+3\,\bigr].
\end{equation}
Hence the pressure-to-energy-density ratio becomes $\frac{P}{\rho}=\frac{nT}{nT\bigl[A+3\bigr]}=\frac{1}{A+3}$. So, starting from the Lorentz-transformed energy density we obtain

\begin{equation}
\frac{\rho'}{\rho}
=\gamma^{2}\!\left(1+\frac{P}{\rho}v^{2}\right)
=\gamma^{2}\!\left(1+\frac{v^{2}}{A+3}\right).
\end{equation}
Defining the effective temperature $T_{\mathrm{eff}}$ by the condition
\(\rho(T_{\mathrm{eff}})=\rho'\) and using the leading-order scaling
\(\rho\propto T^{4}\) (valid for the thermodynamic mapping used here),
we have

\begin{equation}
\left(\frac{T_{\mathrm{eff}}}{T}\right)^{4}=\frac{\rho'}{\rho}
=\gamma^{2}\!\left(1+\frac{v^{2}}{A+3}\right),
\end{equation}
equivalently

\begin{equation}
\frac{T_{\mathrm{eff}}}{T}
=\left[\gamma^{2}\!\left(1+\frac{v^{2}}{A(m/T)+3}\right)\right]^{1/4}.
\label{Teff_final}
\end{equation}
In the ultrarelativistic regime $(T\gg m)$, we have \(\displaystyle m/T\to 0\) and \(K_{1}/K_{2}\to 0\) appropriately so that \(A\to 0\), and (\ref{Teff_final}) reduces to photon-gas result:
\(\;T_{\mathrm{eff}}/T=[\gamma^{2}(1+v^{2}/3)]^{1/4}\;\).
On the other hand in the nonrelativistic limit \(\displaystyle m/T\gg 1\), we have \(K_{1}(x)/K_{2}(x)\to 1\) to leading exponential order, so \(A\sim m/T\gg 1\).
Consequently \(P/\rho\approx 1/(A+3)\ll 1\) and the velocity-dependent enhancement of \(T_{\mathrm{eff}}\) is reduced compared with the photon case. Therefore, both massless and massive gases exhibit the same qualitative trend:
the effective temperature grows with $v$.

It should be emphasized that the assumption $\rho \propto T^{4}$ used in deriving
Eq.~(\ref{Teff_final}) is strictly valid only in the photon gas limit,
where the equation of state satisfies $P=\rho/3$.
For a massive Maxwell--J\"uttner gas the full dependence of $\rho$ on $T$
is given by (\ref{AAA}) and the relation between $\rho$ and $T_{\mathrm{eff}}$ should in principle be obtained
by solving $\rho(T_{\mathrm{eff}})=\gamma^{2}(\rho+Pv^{2})$ explicitly.
Nevertheless, the above simplified form captures the leading behaviour
and provides a continuous interpolation between the ultrarelativistic and
nonrelativistic limits. Expanding Eq.~(\ref{Teff_final}) for small velocities $v\ll1$ gives

\begin{equation}
\frac{T_{\mathrm{eff}}}{T}\simeq 1 + C(A)\,v^{2} + \mathcal{O}(v^{4}),
\qquad
C(A)=\frac{A+4}{4(A+3)}.
\label{eq:Teff_expand}
\end{equation}
Hence $C(0)=1/3$ in the photon limit, while $C\to1/4$ for $A\to\infty$
(corresponding to a completely nonrelativistic gas).
For intermediate values, $A(m/T)\!\sim\!0.3\!-\!5$, the coefficient lies in the range
$C(A)\!\approx\!0.28\text{--}0.33$, showing that the quadratic correction varies only mildly
with the particle mass.

It should also be noted that if $T_{\mathrm{eff}}$ were defined using the
\emph{total} energy density including the rest-mass contribution $nm$,
bulk motion would artificially enhance $\rho'$ by a factor $\gamma^{2}nm$,
producing much larger apparent coefficients of $v^{2}$.
Therefore, throughout this analysis $T_{\mathrm{eff}}$ is understood to be
defined from the internal energy density $\rho-nm$, which yields the physically
meaningful behaviour summarized in Eq.~(\ref{eq:Teff_expand}).

\begin{figure}[h!]
\centering
\includegraphics[width=0.5\textwidth]{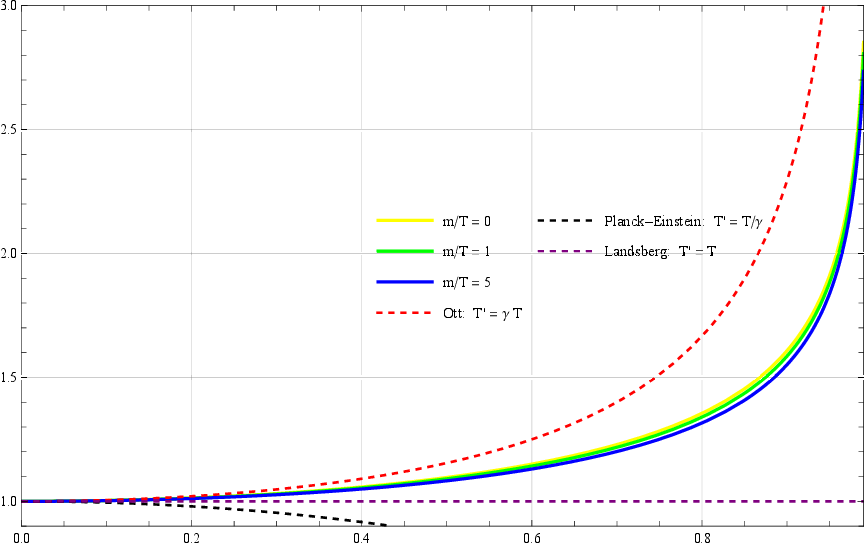}
\caption{Dependence of the effective temperature ratio $T_{\mathrm{eff}}/T$
on the velocity $v$ for several values of the dimensionless
mass-to-temperature ratio $m/T$, obtained from Eq.~(\ref{Teff_final}).
The solid curve corresponds to the photon limit ($m/T\!\to\!0$),
approaching $T_{\mathrm{eff}}\simeq T(1+\tfrac{1}{3}v^{2})$,
while the dashed and dotted curves represent moderately and
completely nonrelativistic gases, for which the asymptotic expansion
yields $T_{\mathrm{eff}}\simeq T(1+\tfrac{1}{4}v^{2})$.
All curves smoothly interpolate between these two limits.}
\label{fig:Teff_MJ}
\end{figure}

The full dependence of $T_{\mathrm{eff}}/T$ on the velocity $v$ for different
mass-to-temperature ratios $m/T$ can be obtained either analytically in the
small-$v$ limit of Eq.~(\ref{eq:Teff_expand}) or numerically by solving
Eq.~(\ref{Teff_final}) with the exact Maxwell--J\"uttner functions $K_{1}$ and $K_{2}$.
Figure~\ref{fig:Teff_MJ} illustrates the resulting curves for several values
of $m/T$.  As expected, the deviation from the photon-gas behaviour becomes
smaller as the gas becomes less relativistic, and the effective temperature
approaches the limiting form $T_{\mathrm{eff}}\simeq T(1+\tfrac{1}{4}v^{2})$
in the nonrelativistic regime.  Conversely, in the ultrarelativistic limit
($m/T\!\to\!0$) the familiar photon-gas dependence
$T_{\mathrm{eff}}\simeq T(1+\tfrac{1}{3}v^{2})$ is recovered.

This behaviour demonstrates that the apparent anisotropy of the energy--momentum
tensor in the moving frame has only a mild influence on the temperature,
and that the functional dependence of $T_{\mathrm{eff}}/T$ on $v$ evolves
smoothly between the two limiting cases.  The present formulation therefore
provides a unified description of relativistic temperature transformation
for both massless and massive gases within the same covariant thermodynamic
framework.

In figure~\ref{fig:Teff_MJ}, in addition of the the behavior of $T_{\mathrm{eff}}/T$ as a function of $v$
for several mass ratios $m/T$, the predictions of the three classical transformation laws are shown. In all cases considered,
the effective temperature increases with motion,
so the sign of the effect supports the Ott–Eddington–Møller interpretation.
However, the precise scaling is not a simple Lorentz factor $\gamma$,
but depends on the internal equation of state of the medium.
This confirms that there is no universal temperature–transformation law:
each physical system defines its own relation
between $T_{\mathrm{eff}}$ and $T$ through its specific $f(T)$ and $g(T)$.

\section{Relativistic Fermi gas (electron gas)}

The formalism developed above can be straightforwardly extended to the relativistic
Fermi systems, such as an electron gas.
The only structural change is that the single-particle occupation numbers obey
Fermi--Dirac statistics and the thermodynamic integrals must be evaluated accordingly.
Below, we summarize the relevant expressions and show how the effective temperature
$T_{\mathrm{eff}}$ is obtained under a Lorentz boost. For a gas of spin-$\tfrac12$ fermions with degeneracy $g$ (for electrons $g=2$)
the number density, pressure and energy density in the local rest frame are

\begin{align}
n(T,\mu) &= \frac{g}{2\pi^{2}}\int_{0}^{\infty}\frac{p^{2}\,dp}{\exp\!\big(\tfrac{E(p)-\mu}{T}\big)+1},
\label{eq:n_FD}\\[4pt]
P(T,\mu) &= \frac{g}{6\pi^{2}}\int_{0}^{\infty}\frac{p^{4}}{E(p)}\,
\frac{dp}{\exp\!\big(\tfrac{E(p)-\mu}{T}\big)+1},
\label{eq:P_FD}\\[4pt]
\rho(T,\mu) &= \frac{g}{2\pi^{2}}\int_{0}^{\infty} E(p)\,p^{2}\,
\frac{dp}{\exp\!\big(\tfrac{E(p)-\mu}{T}\big)+1},
\label{eq:rho_FD}
\end{align}
where \(E(p)=\sqrt{p^{2}+m^{2}}\). These integrals define the equation of state
$\rho=\rho(T,\mu)$ and $P=P(T,\mu)$.

Under a Lorentz boost with speed \(v\) along the \(x\)-axis the energy density
measured by the moving observer is given as before by equation (\ref{eq:Teff_def}). We define the effective temperature \(T_{\mathrm{eff}}\) (for a given chemical
potential in the comoving frame) by the operational condition

\begin{equation}
\rho(T_{\mathrm{eff}},\mu) = \rho'.
\end{equation}
That is, the moving observer infers \(T_{\mathrm{eff}}\) by inverting the
rest-frame relation \(\rho=\rho(T,\mu)\) while keeping \(\mu\) (the comoving
chemical potential) fixed in this operational protocol.\footnote{Here we assume that the chemical potential $\mu$ behaves as a local
thermodynamic scalar. This assumption is consistent with the covariant
equilibrium distribution $f \sim \exp\!\left[-\frac{p_\mu u^\mu - \mu}{T}\right],$
widely used in relativistic kinetic theory, where $p_\mu u^\mu$ is a
Lorentz scalar. Under this assumption the equilibrium form of the
distribution is preserved in different inertial frames, allowing a
consistent operational comparison of temperatures extracted from the
same equation of state.}

It is convenient to introduce the dimensionless ratios \(x\equiv m/T\)
and \(\alpha\equiv\mu/T\). Define the function

\begin{equation}
A_F\!\bigl(x,\alpha\bigr)\;\equiv\;3\frac{\rho-3P}{3P}
\;=\;\frac{\rho}{P}-3,
\label{eq:A_F_def}
\end{equation}
which generalizes the Maxwell--J\"uttner $A(m/T)$ to the Fermi case.
Using \(\rho'=\gamma^{2}(\rho+Pv^{2})\) we obtain the compact relation

\begin{equation}\label{Tem. Fermi}
\left(\frac{T_{\mathrm{eff}}}{T}\right)^{4}
= \gamma^{2}\!\left(1+\frac{v^{2}}{A_F(x,\alpha)+3}\right),
\end{equation}
which is formally identical to the Maxwell--J\"uttner case with \(A\mapsto A_F\).
Expanding for small velocities \(v\ll1\) gives

\begin{equation}
\frac{T_{\mathrm{eff}}}{T}\simeq 1 + C_F(x,\alpha)\,v^{2} + \mathcal{O}(v^{4}),
\qquad
C_F(x,\alpha)=\frac{A_F(x,\alpha)+4}{4\big(A_F(x,\alpha)+3\big)}.
\end{equation}
Thus the quadratic coefficient is determined entirely by the equilibrium
thermodynamic ratio \(P/\rho\) (or equivalently \(A_F\)) evaluated at \((T,\mu)\).

In the nondegenerate regime for which \(\alpha=\mu/T\ll -1\), the Fermi--Dirac integrals reduce to the Boltzmann
(Maxwell--J\"uttner) form and \(A_F(x,\alpha)\to A(x)\), recovering the
results of Section~5. In particular for \(x\ll1\) (ultrarelativistic)
one finds \(C_F\to 1/3\). On the other hand, in the strongly degenerate regime $T\ll \mu$ (cold Fermi gas), the thermodynamics is dominated by states
up to the Fermi momentum \(p_F\) defined by \(\mu=\sqrt{p_F^2+m^2}\).
At \(T=0\), the integrals have closed forms

\begin{align}
n &= \frac{g}{6\pi^{2}}p_F^3, \\
\rho &= \frac{g}{8\pi^{2}}\left[
p_F\mu\,(2p_F^2+\mu^2)-m^4\ln\!\Big(\frac{p_F+\mu}{m}\Big)\right], \\
P &= \frac{g}{24\pi^{2}}\left[
p_F\mu\,(2p_F^2-3m^2)+3m^4\ln\!\Big(\frac{p_F+\mu}{m}\Big)\right].
\end{align}
From these, one computes \(A_F(x,\alpha)\) in the degenerate limit and hence
the small-$v$ coefficient \(C_F\). Notably, in the strong degenerate limit
the coefficient \(C_F\) tends to a value bounded between \(1/4\) and \(1/3\)
depending on the degree of relativity of the Fermi sea (ultrarelativistic
degenerate gas approaches \(1/3\), while nonrelativistic degenerate gas is
closer to \(1/4\)).

For interpreting $T_{\mathrm{eff}}$ in the Fermi case two remarks are important: (i) \emph{Chemical potential:} In our operational definition we hold the
comoving chemical potential \(\mu\) fixed while extracting
\(T_{\mathrm{eff}}\) from \(\rho\). This reflects the physical
protocol where the moving observer infers a temperature assuming
thermal and chemical properties of the comoving fluid are unchanged.
Alternative protocols (e.g. fixing particle density \(n\) instead)
lead to different numerical values of \(T_{\mathrm{eff}}\) and should
be stated explicitly if used. (ii) \emph{Degeneracy vs. thermal effects:} In degenerate systems the
thermodynamic temperature plays a diminished role compared to the
Fermi energy; therefore \(T_{\mathrm{eff}}\) extracted from \(\rho\)
may be less informative physically. For such systems it can be more
illuminating to consider an effective thermal excitation energy
or to compare the boost-induced change in internal energy per particle.

Evaluation of \(A_F(x,\alpha)\) requires numerical quadrature of the
Fermi--Dirac integrals (Eqs.~\ref{eq:n_FD}--\ref{eq:rho_FD}) for the desired
pair \((x,\alpha)\). Once \(A_F\) is obtained, \(T_{\mathrm{eff}}\) follows
from the algebraic inversion of the relation (\ref{Tem. Fermi}).
For practical purposes one may tabulate \(A_F\) on a grid in \((x,\alpha)\)
and interpolate; this provides an efficient way to construct
plots analogous to figure~\ref{fig:Teff_MJ} for degenerate or partially
degenerate electron gases.

\begin{figure}[h!]
\centering
\includegraphics[width=0.5\textwidth]{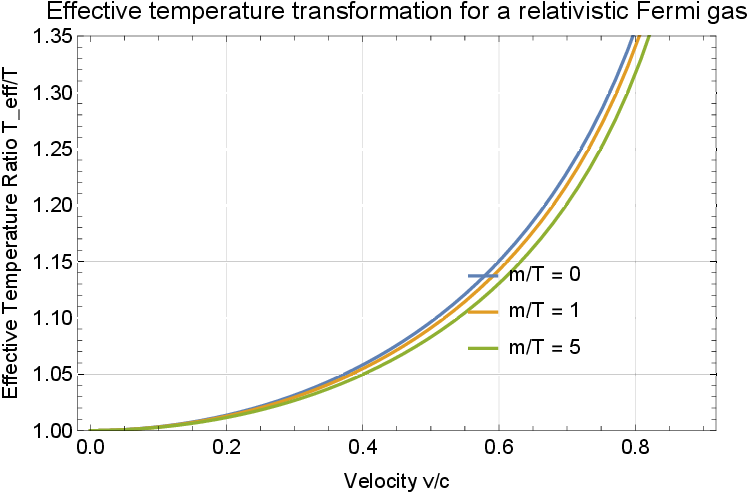}
\caption{Effective temperature ratio $T_{\mathrm{eff}}/T$ versus observer velocity $v/c$
for representative values of the dimensionless parameter $A_F$ (approximating
$m/T$): $A_F=0$ ($m/T\approx0$), $A_F=1$ ($m/T\sim1$) and $A_F=5$ ($m/T\gg1$).
Curves are obtained from the relation
$\;T_{\mathrm{eff}}/T=[\gamma^{2}(1+v^{2}/(A_F+3))]^{1/4}\;$ and show the smooth
interpolation between the ultrarelativistic ($T_{\mathrm{eff}}\simeq T(1+\tfrac{1}{3}v^{2})$)
and nonrelativistic ($T_{\mathrm{eff}}\simeq T(1+\tfrac{1}{4}v^{2})$) limits.}
\label{fig:Teff_Fermi}
\end{figure}
As illustrated in figure~\ref{fig:Teff_Fermi}, the ratio 
$T_{\mathrm{eff}}/T$ exhibits a monotonically increasing trend
with the observer’s velocity $v$, for all considered values of
the mass-to-temperature ratio $m/T$.
The dependence becomes more pronounced in the ultrarelativistic
limit (smaller $m/T$), where the gas behaves nearly photon-like,
and the effective temperature follows the approximate expansion
$T_{\mathrm{eff}} \simeq T(1+\tfrac{1}{3}v^{2})$.
As the gas becomes more nonrelativistic (larger $m/T$), the
velocity dependence weakens and gradually approaches the
nonrelativistic limit $T_{\mathrm{eff}} \simeq T(1+\tfrac{1}{4}v^{2})$.
This smooth interpolation between the two regimes indicates that
the covariant prescription based on the transformed
energy–momentum tensor provides a consistent and unified picture
of temperature transformation for both Maxwell–Boltzmann and
Fermi–Dirac statistics.

The covariant formulation developed above for the relativistic
Fermi gas can be naturally extended to other quantum gases.
In particular, for bosonic systems one simply replaces the
Fermi–Dirac occupation factor
$[\,\exp((E-\mu)/T)+1\,]^{-1}$ by the Bose–Einstein form
$[\,\exp((E-\mu)/T)-1\,]^{-1}$ in the thermodynamic integrals,
leading to an analogous function $A_B(m/T,\mu/T)$ and hence
to the same structural relation for the effective temperature,
$T_{\mathrm{eff}}/T=[\,\gamma^2(1+v^2/(A_B+3))\,]^{1/4}$.
More generally, the same reasoning can be applied to multi-component
plasmas or mixtures of fermionic and bosonic species, each
characterized by its own rest mass and chemical potential.
In all such cases, the essential covariant structure of the
temperature transformation, derived from the boosted energy–momentum
tensor, remains valid.

\section{Discussion}

The three physical systems analyzed above---blackbody radiation, the relativistic ideal gas and the Fermi electron gas---
lead to a consistent qualitative picture:
the temperature inferred by a moving observer from the total energy density
is \emph{higher} than the rest-frame temperature.
This behavior agrees in sign with the Ott–Eddington–Møller transformation law,
although the quantitative dependence on the velocity differs
from the simple linear factor $\gamma$. Our results above, show that $T_{\mathrm{eff}}$ depends explicitly on the equation of state functions
$f(T)$ and $g(T)$. Consequently, the temperature transformation cannot be universal which means that 
each physical system has its own characteristic response to motion,
governed by its microscopic statistics.
In the ultrarelativistic limit, all systems approach the photon–gas form,
while in the nonrelativistic regime the enhancement of $T_{\mathrm{eff}}$
with $v^{2}$ becomes even stronger. This system-dependence implies that temperature is not a fundamental Lorentz scalar,
but a macroscopic quantity whose transformation
reflects the underlying energy–momentum distribution of the medium. In contrast, note that the fundamental
thermodynamic laws remain fully covariant, since they are formulated in terms
of the stress–energy tensor $T^{\mu\nu}$ and the entropy current $s^\mu$,
whose transformation properties are universal and independent of the specific
fluid. Therefore, relativistic thermodynamics does not require a universal
temperature transformation law; rather, temperature is an observer–dependent
derived quantity whose transformation reflects the microscopic properties of
the medium under consideration.

A more general covariant framework employs the
inverse-temperature four-vector $\beta^{\mu}$,
defined as

\begin{equation}
\beta^{\mu} = \frac{u^{\mu}}{k_{B}T},
\label{eq:betavector}
\end{equation}
where $u^{\mu}$ is the four-velocity of the system.
In this approach, the temperature $T$ is invariant,
while the directional effects of motion are encoded in $\beta^{\mu}$.
The local equilibrium distribution in kinetic theory
takes the manifestly covariant form

\begin{equation}
f(x,p) \propto \exp(-\beta_{\mu} p^{\mu}),
\label{eq:covariant_distribution}
\end{equation}
which ensures the Lorentz invariance of the entropy four-current
and the energy–momentum tensor.
The ``temperature'' measured by a specific observer
depends on the scalar contraction $\beta_{\mu} w^{\mu}$,
where $w^{\mu}$ is the observer's four-velocity.
This leads naturally to observer-dependent effective temperatures,
reconciling the apparent conflict among the classical proposals.

From the present analysis, the classical transformation laws
can be interpreted as corresponding to different
choices of what quantity is held invariant under Lorentz boosts. The Planck–Einstein law ($T' = T/\gamma$) assumes the invariance of entropy and total heat $Q$, but does not preserve the covariant form of the first law. The Ott–Eddington–Møller law ($T' = \gamma T$) follows from treating heat as part of the energy–momentum four-vector and is consistent with the transformation of energy flux. Finally, the Landsberg law ($T'=T$) corresponds to the invariant scalar interpretation adopted in covariant statistical mechanics through Eq.~(\ref{eq:betavector}).
Our results suggest that the effective temperature derived
from measurable energy densities behaves according to the Ott interpretation,
while the invariant formulation (\ref{eq:betavector})
provides the deeper covariant meaning of temperature in relativity.

The coexistence of these two perspectives---operational (Ott-like) and covariant (Landsberg-like)---
reflects the dual character of temperature in relativistic physics.
As an intensive variable derived from entropy and energy,
temperature behaves as a scalar parameter in the equilibrium distribution (\ref{eq:covariant_distribution});
but as an observable inferred from energy density or radiation intensity,
it acquires observer dependence through Lorentz transformations.

In practical terms, this means that statements such as
``a moving body appears hotter'' or ``cooler''
have no absolute meaning unless the measurement protocol is specified.
Different operational definitions (thermometric, calorimetric or spectroscopic)
lead to different transformation laws.

It is worth emphasizing that the present analysis should not be confused
with the Unruh effect. Here all observers are inertial and the thermal
system already possesses a physical equilibrium temperature in its
comoving frame. The problem addressed in this work is therefore the
Lorentz transformation of temperature between inertial observers, rather
than the emergence of thermal behavior due to acceleration.

The present analysis is also formulated entirely in flat Minkowski
spacetime, where $g_{00}=1$. Therefore, the Tolman--Ehrenfest relation $T\sqrt{g_{00}}=\mathrm{const}$
reduces trivially to a constant equilibrium temperature. In this sense,
our results neither modify nor contradict the standard gravitational
equilibrium condition. Rather, they concern the Lorentz transformation
of thermodynamic quantities between inertial observers in special
relativity. We expect that an extension of the present framework to
curved spacetime should recover the Tolman relation locally, at least
in the weak--field limit.

\section{Summary}
In this work we have examined the relativistic transformation of temperature
from a microscopic and covariant perspective.
Rather than assuming any specific transformation law \emph{a priori},
we started from the general energy–momentum tensor of an isotropic system
and derived the effective temperature that a moving observer
would infer from the transformed energy density.
This approach allowed a direct comparison between
theoretical proposals and physically measurable quantities. our main conclusions are:

${\bullet}$ For all three models considered in this article: the photon gas, the relativistic ideal gas
described by the Maxwell–Jüttner distribution and the electron gas described by the Fermi-Dirac distribution,
the effective temperature \(T_{\mathrm{eff}}\)
increases monotonically with the velocity of the observer.

${\bullet}$ The sign of this variation agrees with the
Ott–Eddington–Møller transformation (\(T' = \gamma T\)),
whereas the magnitude depends on the microscopic equation of state.
For small velocities, we found approximately

\begin{equation}
\frac{T_{\mathrm{eff}}}{T} \simeq 1 + C(m/T) v^{2},
\end{equation}
where \(C \approx 0.33\) for ultrarelativistic systems
and \(C \approx 0.25\) in the nonrelativistic regime.

${\bullet}$ The Planck–Einstein law (\(T' = T/\gamma\))
predicts the opposite behavior and is inconsistent
with the Lorentz transformation of the energy–momentum tensor,
while the Landsberg invariant form (\(T'=T\))
captures only the covariant statistical meaning of temperature,
not its operational value.

${\bullet}$ The absence of a universal transformation rule
reflects the fact that ``temperature'' is not a fundamental
Lorentz scalar but an observer-dependent parameter,
emerging from specific measurement definitions.

${\bullet}$ The results support a unified interpretation in which
temperature enters the relativistic formalism through
the inverse-temperature four-vector $\beta^{\mu} = \frac{u^{\mu}}{k_{B}T}$. In this representation, \(T\) itself remains invariant,
while the contraction \(\beta_{\mu}w^{\mu}\)
determines the temperature measured by an observer
with four-velocity \(w^{\mu}\).
This picture reconciles the invariant and operational viewpoints:
the Ott transformation arises from the observer dependence
of the scalar product \(\beta_{\mu}w^{\mu}\),
while the Landsberg formulation maintains covariance at the microscopic level.

In summary, our results indicate that the apparent temperature of a moving system
depends not only on the relative velocity but also on its microscopic constitution.
A covariant statistical treatment, based on the inverse-temperature four-vector,
provides the natural framework for understanding this dependence.
The century-long debate over relativistic temperature transformation
thus may find a coherent resolution:
different classical laws describe different operational definitions of temperature,
all consistent within the broader covariant thermodynamic picture.
\vspace{5mm}\newline \noindent {\bf
Acknowledgement}\vspace{2mm}\noindent\newline
The authors would like to thank Kourosh Nozari for a
careful reading of the manuscript and very useful comments.

\end{document}